\documentclass[sigconf]{acmart}

\AtBeginDocument{%
  \providecommand\BibTeX{{%
    \normalfont B\kern-0.5em{\scshape i\kern-0.25em b}\kern-0.8em\TeX}}}

\begin{document}


\title{Learning from Teaching Assistants to Program with Subgoals: Exploring the Potential for AI Teaching Assistants}

\author{Changyoon Lee}
\orcid{0009-0007-3781-5296}
\email{changyoon.lee@kaist.ac.kr}
\affiliation{%
  \institution{School of Computing, KAIST}
  \city{Daejeon}
  \country{South Korea}
}

\author{Junho Myung}
\authornote{Authors contributed equally as second authors to this research.}
\email{junho00211@kaist.ac.kr}
\affiliation{%
  \institution{School of Computing, KAIST}
  \city{Daejeon}
  \country{South Korea}
}

\author{Jieun Han}
\email{jieun_han@kaist.ac.kr}
\authornotemark[1]
\affiliation{%
  \institution{School of Computing, KAIST}
  \city{Daejeon}
  \country{South Korea}
}

\author{Jiho Jin}
\email{jinjh0123@kaist.ac.kr}
\authornotemark[1]
\affiliation{%
  \institution{School of Computing, KAIST}
  \city{Daejeon}
  \country{South Korea}
}

\author{Alice Oh}
\email{alice.oh@kaist.edu}
\affiliation{%
  \institution{School of Computing, KAIST}
  \city{Daejeon}
  \country{South Korea}
}


\begin{abstract}

With recent advances in generative AI, conversational models like ChatGPT have become feasible candidates for TAs. 
We investigate the practicality of using generative AI as TAs in introductory programming education by examining novice learners' interaction with TAs in a subgoal learning environment. 
To compare the learners' interaction and perception of the AI and human TAs, we conducted a between-subject study with 20 novice programming learners. Learners solve programming tasks by producing subgoals and subsolutions with the guidance of a TA. 
Our study shows that learners can solve tasks faster with comparable scores with AI TAs.
Learners' perception of the AI TA is on par with that of human TAs in terms of speed and comprehensiveness of the replies and helpfulness, difficulty, and satisfaction of the conversation.
Finally, we suggest guidelines to better design and utilize generative AI as TAs in programming education from the result of our chat log analysis.



\end{abstract}



\begin{teaserfigure}
  \includegraphics[width=\textwidth]{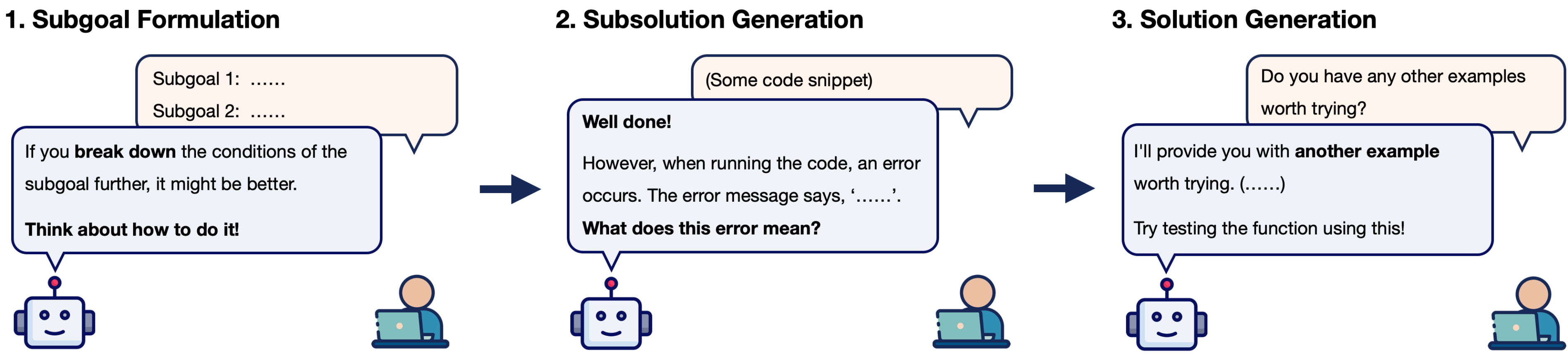}
  \caption{The learning workflow the learners go through with TAs consists of three steps: 1) Subgoal Formulation, 2) Subproblem Generation, and 3) Solution Generation.}
  \Description{Learner interacting with the TA when solving programming tasks. The above texts are taken from the real interaction of the users in our user study.}
  \label{fig:teaser}
\end{teaserfigure}


\maketitle

\section{Introduction}
A common goal of introductory programming courses is to teach learners how to program. While writing code is an essential requirement of programming, students are expected to learn other important concepts such as debugging, designing algorithms, and techniques in programming~\cite{Becker19}. Learners also learn computational thinking~\cite{wing2006computational}, a new way of thinking that involves the abstraction of problems, decomposing them, and re-composing them into working solutions~\cite{mccracken2001multi,agbo2019systematic}. Having to learn these new concepts in a single course presents difficulties to the learners~\cite{konecki2014problems}, and if this difficulty is not appropriately alleviated, the learners may lose motivation and even drop out of the course~\cite{rountree2002success,kinnunen2006dropout}.

Therefore, many programming courses employ teaching assistants (TAs) to closely attend to the learners' needs and provide feedback. TAs play a crucial role in correcting learners' misconceptions and fixing errors in their code, thereby enhancing their overall learning gain~\cite{corbett2001locus,gusukuma2018misconception,marwan2020adaptive}. Having a sufficient number of TAs allows learners to receive individual care by getting help in solving programming tasks and clarifying programming concepts~\cite{mirza2019undergraduate,riese2021challenges}.

With recent advances in generative AI and Large Language Models (LLMs), the educational field has discovered some exciting opportunities for assisting learners. Open-domain generative models are trained to be empathetic~\cite{roller-etal-2021-recipes} and perform well in unseen contexts for better reliability~\cite{brown2020language}, which are important qualities for TAs when interacting with learners. More recent large generative models such as ChatGPT\footnote{https://chat.openai.com/}, LLaMA~\cite{touvron2023llama}, and Bard\footnote{https://bard.google.com/} can also participate in longer conversations by remembering the context of the conversations.

In the context of programming education, these models also show a remarkable ability to understand, generate, and explain code, making them strong candidates for TAs in programming courses~\cite{savelka2023thrilled}. They can fix errors present in the code, explain why the errors occur, and discuss possible approaches to solve various programming tasks. AI coding assistants have been shown to relieve the cognitive load and struggles of learners, allowing them to perform better and faster in solving programming tasks~\cite{kazemitabaar2023studying}. 

However, prior work has explored the pedagogical abilities of conversational agents and showed that they still fall behind human teachers in providing necessary help to the students~\cite{tack2022teacher}. Also, generative models sometimes produce inconsistent code, generating different responses to the same prompt on different occasions~\cite{jonsson2022cracking}. This underscores the importance of a meticulous approach when implementing LLMs within real classrooms.

In this note, we introduce the concept of subgoal learning to novice learners when conversing with the AI TAs to learn programming concepts. Subgoal learning is well known to be an effective learning strategy in the STEM domain by helping students break down complex problems into smaller counterparts \cite{Catrambone90}. We investigate the potential of AI TAs in educating novice learners in a structural manner by comparing their interactions to those of human TAs.

To explore the feasibility of employing generative AI as TAs in introductory programming courses and study how learners and AI TAs interact, we conducted a between-subject study with 20 novice programming learners, of whom 10 had no previous experience in programming, and the remaining 10 had only taken an introductory programming course before. Half of the learners in each group solved 4 programming tasks in each of the two sessions separated by a week with the aid of an AI TA powered by ChatGPT while the other half solved with the aid of a human TA. The learners followed our learning workflow of dividing the task into subgoals and solving them by writing subsolutions, through which learners are expected to develop computational thinking skills. An assessment of AI's ability to help plan an algorithm and write the code for it within our learning workflow was conducted. Learners participated in a survey after each programming session that asked for their perceptions about their satisfaction with the conversation with the TA and the learning workflow. Seven learners participated in a retention test and an interview conducted a week after the second programming session. 

Our results show that learners who received help from the AI TA solved the programming tasks faster and attempted more tasks. They also achieved comparable scores for the tasks. In the survey, learners reported that AI TA's replies were prompt, sufficiently detailed, and helpful throughout the workflow. Moreover, learners were satisfied with the conversation with the AI and perceived that it was generally uncomplicated and helpful for learning programming. The analysis of the chat log reveals the different behaviors of the AI and human TAs and design opportunities for a better AI TA.

The contributions of this paper are as follows:
\begin{itemize}
     \item Results from a between-subject study that shows the learners' programming performance with human and AI TAs.
     \item An analysis of the strengths and weaknesses of AI TA in programming from the chat log analysis
     \item A set of design guidelines for designing an effective AI TA in programming
\end{itemize}

\section{Related Work}
\subsection{Subgoal Learning}
Subgoal learning is a method designed to assist students in breaking down complex problem-solving procedures into smaller structural components within the STEM domain \cite{Catrambone90}. This strategy has demonstrated significant effectiveness in improving learners' ability to transfer knowledge across tasks that share similar subgoals \cite{Margulieux12, Catrambone98}.
In the context of programming education, subgoal learning is known to be helpful in reducing the extraneous cognitive load of the learners, thereby enhancing their problem-solving performance \cite{Margulieux12,  Atkinson00}. 

The effectiveness of subgoal learning is further amplified when implemented as an active learning strategy. Conventional subgoal learning involved the provision of pre-defined subgoal labels, making learners learn in a passive manner \cite{Morrison15}. Such a passive learning approach was found to be less effective compared to self-directed learning methods, which involve self-reflection and explanation of the hierarchical structure of the solutions \cite{Morrison15, lauren19}. Yet, proper guidance or feedback is necessary to correct learners' misconceptions of the concept when creating subgoals by themselves \cite{lauren19, jin19}.

Previous studies on integrating self-motivated subgoal learning into the programming domain focus on utilizing learners' experience as guidance to other novice learners through interactive platforms.
Crowdy introduces the concept of \textit{learnersourcing} to generate subgoal labels from introductory web programming videos, thereby assisting learners in more effective comprehension of programming concepts \cite{Weir15}. 
Algosolve supports novice programming learners in creating high-quality subgoal labels by leveraging peer examples \cite{choi22}. 
However, there is still a lack of research done on utilizing AI models as an instructor to clarify learners' questions and guide them in the process of self-labeling subgoal tasks. 

\subsection{Generative AI for Programming Education}
Generative AI exhibits remarkable performance in various programming tasks, such as code repair \cite{surameery23, xia23}, code summarization \cite{ahmed22}, code generation \cite{chakraborty22, lu21}, and even code explanation \cite{chen2023gptutor}. This recent advancement in generative AI opens up numerous opportunities to support programming education for various stakeholders, including instructors, students, and teaching assistants.

\subsubsection{Instructors}
Leveraging LLMs can significantly elevate the effectiveness of instructors in programming courses. These models serve as valuable tools for addressing various instructional tasks, such as answering students' questions, assessing assignments and quizzes, and developing lesson plans \cite{rahman23}. Codex \cite{chen21}, a descendant of OpenAI GPT-3 optimized for programming, serves as an effective aid for university instructors by automating the generation of programming exercises, tests, and solutions, thus saving time and enhancing the quality of educational materials \cite{Sarsa22}. Instructors can also harness LLMs to deliver personalized learning support to their students, for instance by identifying specific areas where students are struggling with \cite{rahman23}. 

\subsubsection{Students}
LLMs can also enhance students' programming learning experience. Codex performs better on answering most introductory programming problems than average students \cite{finnie22}. Therefore, novice learners can gain a deeper understanding of basic programming concepts with line-by-line code explanations generated by LLMs \cite{MacNeil23}. They can also receive feedback and detect bugs before they submit their assignments for grading \cite{Sarsa22}. The nearly instantaneous provision of feedback and explanations makes generative AIs more accessible and convenient for learners compared to traditional human instructors.


\subsection{Generative AI as TA}
TAs in programming education are generally more approachable to the students while being skilled at identifying and addressing each student's special needs \cite{Riese20}. However, TAs face challenges of managing multiple students simultaneously \cite{riese2021challenges, McDonald23}. Students often possess varying levels of knowledge and different needs, making it even more demanding for TAs to provide personalized feedback effectively \cite{riese2021challenges}. Because of this, LLMs can be utilized to offer personalized and immediate responses \cite{zhai22}, just as TAs are supposed to do.

Nevertheless, the straightforward integration of LLMs in this role may lead to various challenges. Despite GPT-3's strong performance on conversation uptake, it falls behind in pedagogical ability when compared to human teachers \cite{tack2022teacher}. Additionally, an over-reliance on generative AI for generating answers could potentially hinder the development of learners' critical thinking and problem-solving skills \cite{rahman23}. These limitations underscore the importance of strategic planning when implementing LLMs in the classroom.

Yet, to the best of our knowledge, examining the performance of LLM compared to human TA has been underexplored. Moreover, the existing literature scarcely explores how to optimally leverage LLMs as TAs, especially in introductory programming education settings.

\section{Learning Workflow Design}
With the advances in generative AI technology, programmers can receive significant help when writing the code. The same is true for programming learners; generative AI can write a code outline or even generate the solution code snippet for a given task when only given natural language instructions. This makes learning to write code a less crucial part of programming education as the generative AI can help to write most of the code. On the other hand, learning computational thinking, designing the solution, and understanding and debugging code remain important for a programmer. 

Our learning workflow is designed to focus on helping learners practice computational thinking and planning out the solution to a problem with the help of generative AI or a human teaching assistant. Learners are guided through a series of steps that help them think about how to divide the task into smaller problems and solve them. In each step, learners come up with mini-solutions that are used in the following step and lead to the final solution. Throughout the learning workflow, we observe how students interact and communicate with a generative AI-powered TA compared to a human TA and how feasible using generative AI as a TA for CS education is. We describe the workflow and steps in detail.

\subsection{Overall Learning Workflow}
In our learning workflow, learners solve each programming task as they progress through three steps: 1) \textbf{Subgoal Formulation step}, where learners break down the task into smaller and more manageable problems, 2) \textbf{Subsolution Generation step}, where learners tackle each of the subgoals they have formulated and implement a solution for them, and 3) \textbf{Solution Generation step} where learners combine their subsolutions into a single code that solves the programming task. The subgoal formulation step drives learners to understand and organize the task and devise a plan for the final solution. The subsolution generation step helps learners focus on a subgoal at a time and progressively write the program. The solution generation step allows learners to review their subsolutions and debug them.

\begin{figure*}[]
    \centering
    \includegraphics[width=\textwidth]{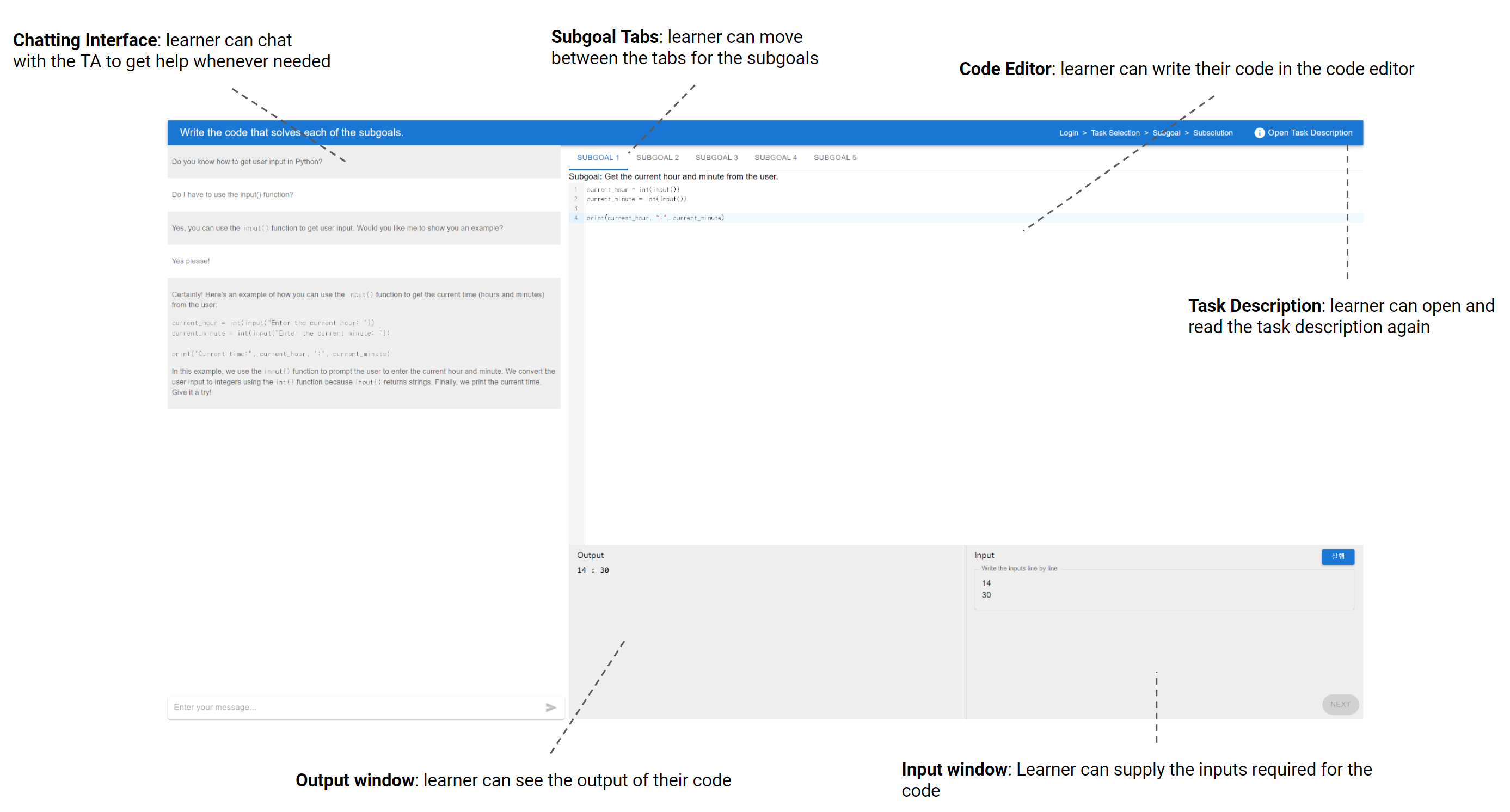}
    \caption{System for solving the programming tasks using the learning workflow.}
    \label{fig:system}
\end{figure*}

Breaking down the problem into subgoals and writing the solution code can be challenging for the learners, especially if they are new to programming. Therefore, at every step of the workflow, a TA, either an instance of prompted generative AI or a human TA, is available for the learners to ask questions. The TA is told about the steps that the learners go through and is asked to guide learners through the steps by answering their questions and giving feedback. Figure~\ref{fig:system} shows the environment that the learner uses to solve the programming tasks. The learner can chat with the TA, navigate between the subgoals, check the task description, edit the code, and run the code to check the outputs on the system.

\subsection{Subgoal Formulation}
When a learner selects a programming task to solve, the learner proceeds to the subgoal formulation step. The learner is initially provided with the task description that includes the requirements of the task, sample inputs, and sample outputs. The code editor shows the skeleton code for the problem. It remains uneditable for this step to help the learner focus on formulating the subgoals while giving the learner an overview of the code. The chat interface on the system allows the learner to communicate with either an AI or a human TA in real time. The learner engages in a conversation with the TA to produce a set of subgoals for the programming task. Once the learner is satisfied with the set of subgoals, the learner moves on to the next step.

This step is designed to encourage the learner to plan ahead instead of jumping right into coding to solve the programming task. Chatting with the TA gives a chance to the learner to clarify any requirements or misunderstandings they might have about the task. The TAs are instructed to remind the learner about the task description and encourage them to write specific subgoals for the task. In order for the learner to take the initiative in formulating the subgoals, the TAs are also instructed not to provide the subgoals the learner has not already mentioned and only correct the learner if the subgoals seem incorrect. As the learner formulates the subgoals on their own, we expect them to familiarize themselves with the programming task and form an idea of what to expect in the following steps.

\subsection{Subsolution Generation}
After the learner formulates a set of subgoals to the task, they move on to the subsolution generation step. They are presented with the subgoals they have formulated in the previous step, and the system asks them to write the solution code for each of the subgoals. The system provides a code editor that the learner writes their subsolutions on. The code can be executed on the page itself, with input and output features to test their subsolutions as they write. The learner thus can easily check the outcome of their code without leaving the system.
The learner communicates with the TA in the same chat interface as the previous step in real time to get help on how to implement the code that solves each of the subgoals they have formulated. For the generative AI TA, a separate chat session is made for each subgoal and subsolution to isolate each subgoal context for better AI performance.

The learner translates their subgoals into code implementation that solves them in this step. Although generative AI can provide significant assistance when writing code, learning to write code is still an important part of computer science education. The learner communicates with the TA to ask about how to approach the problem algorithmically, how to use a certain syntax or debug the code they are writing. The TAs are instructed to help the learner step by step to create a program solution for a subgoal. The TAs should not provide the solution directly and refrain from providing the solution code to the learner. Only when the learner is struggling the TAs help them by providing some code snippets that can help the learner progress in the task. The learner practices writing code and reading and understanding the code that the TA presents as an example. 

\subsection{Solution Generation}
In the final step, the learner revises the code that they have written in the subsolution generation stage. The user interface of the system is identical to that in the subsolution generation stage. The learner is presented with a new chat session and a code editor that contains all the code that the learner has written in the previous step. 
The learner takes another look at the code and tests the code by running some sample inputs and checking that the code's output is the same as the expected output. For instance, the learner can ask for more sample inputs to test their code, verify their solution with the TA, and modify the code as needed.

The solution generation step is designed to give the learner an opportunity to organize the code they have written into a coherent program. Additionally, if the TA helped the learner write parts of the code in the subsolution generation step, the learner can review the code and understand how the subsolution code works and what role it plays in the overall program. The TAs are instructed to help the learner combine the subsolutions into a final solution. They are also reminded not to provide the solution directly and to provide code only when necessary, as in the previous step. The learner finishes up the implementation in this step and submits the solution.

\section{User Study}
We evaluate the feasibility of using an AI TA in an educational setting. We observe learners' perceptions of the AI TA and how they interact to solve the programming tasks given. The goals of our user study were (1) to assess the feasibility of using a generative AI-powered TA, (2) to observe and analyze how learners interact and perceive a generative AI-powered TA, and (3) to examine the strengths and weaknesses of a generative AI-powered TA in CS education when compared to a human TA.

We conducted a between-subjects study where a participant solves programming tasks using our learning workflow by communicating with either a generative AI-powered TA or a human TA in two sessions, separated by approximately a week. After the two sessions, we interviewed 7 participants to ask about their learning experience and gauge how much they remembered about the content they learned.



\subsection{Participants}
We recruited 20 participants (7 male, 13 female, mean age 23.85, stdev=2.16, max=26, min=19) in Korea who have little to no experience in programming. Our recruitment process involved posting in two online university communities to participate in our two-session study. They reported their perceived proficiency in Python, the programming language used to solve programming tasks in our study, on a 5-point Likert scale. Among the 20 participants, 10 participants had a self-reported proficiency of 1 and had not taken any computer science course. The remaining 10 participants had a self-reported proficiency of 2 and had taken only introductory courses in computer science. 

We randomly assigned 5 participants in each proficiency level to solve the programming tasks with the help of a generative AI-powered TA and assigned the remaining 5 participants to solve with the help of a human TA. Out of all participants, 19 participants completed both sessions. One participant who was assigned to a generative AI-powered TA and had a self-reported proficiency of 2 dropped out of the study and only participated in the first session.
Each session lasted for a maximum of 3 hours, and participants were paid 40,000 KRW (\textasciitilde USD 30) per session. Willing participants were interviewed for an hour and were paid an additional 20,000 KRW (\textasciitilde USD 15). 

\subsection{Generative AI-powered TA}
We used ChatGPT (gpt-3.5-turbo) as the model behind the AI TA since GPT-3.5 performs better on programming when prompted in non-English language compared to Codex \cite{hellas23}. The default temperature of 1.0 was used. All participants prompted the AI TA in Korean, which was their native language. For the subgoal formulation step, the model was prompted to help the learners create subgoals by first reminding them about the task, providing hints, reiterating the subgoals, and giving feedback. The task description was provided to the model.
For the subsolution generation step, the model was prompted to help the learners write a program to solve a subgoal by helping the learners debug, answering questions, likely regarding syntax or function usage, and giving feedback. The subgoal was provided to the model.
For the solution generation step, the model was prompted to help the learners combine the subsolutions to create a program by debugging and correcting any mistakes. All the subgoals were provided to the model. 
The learner's code was appended to each of the learner's messages to provide a full context of the learner's status to the AI TA.

In every step, the model was prompted not to provide direct answers, such as subgoals or code snippets, to the learners. Providing the answers to the learners could have a negative impact on learning, as learners may be tempted to merely copy the provided solution, which would remove the opportunity for the learners to practice producing the answers by themselves and learn on their own. The exact prompts given to the model are reported in Appendix~\ref{appendix:prompts}.


\subsection{Programming Tasks}
Participants were given 4 programming tasks to solve in each session. Programming tasks of similar difficulties were presented in the two sessions. Different sets of programming tasks were selected for the participants with different proficiency and experience in programming. For the participants with a reported proficiency of 1, easier problems that test programming concepts such as input, output, loops, if statements, strings, and lists were given. For the participants with a reported proficiency of 2, more demanding problems that test programming concepts such as sorting, searching, replacing, and indexing were given. 
The difficulty of the tasks was extracted from a crowdsourced platform that collects the difficulty rating of programming tasks~\footnote{https://solved.ac/} when possible; otherwise, the authors who have TA experience in introductory programming courses gauged the difficulty.

The programming tasks were chosen to present enough challenge to the learners while not being too difficult for novices which can make them give up. The tasks' contents span basic programming concepts necessary for programmers to learn and can comprehensively assess the learners' computer science knowledge. We provide the full description of the tasks in Appendix~\ref{appendix:tasks}.






\subsection{Instructions}
Participants were first given instructions on how to use our online system and were told whether they would be solving the programming tasks with a generative AI-powered TA or a human TA. They were also given some examples of subgoals to help them generate their own in the main study. Once they accessed the system, they were presented with four programming tasks and asked to solve them using our learning workflow as much as possible within the time limit of 3 hours. After the programming session, participants answered an online survey where they reported about their experience with the TA and the system. 

\subsection{Measurement}

\subsubsection{Qualitative Measurements}
Learners were asked to rate their experience with the TA and the helpfulness of the learning workflow and the system in the online survey. All of the survey questions were rated using a 7-point Likert scale (1: strongly disagree, 7: strongly agree) followed by an open-ended question asking the reason behind the choice. 

To measure the learners' satisfaction with the experience with the TA, we asked the learners how prompt and detailed the TA's responses were, how useful the TA's responses were for each of the three steps in the learning workflow, how much difficulty they faced when talking to the TA, how helpful the TA was, and how satisfied they were with the communication with the TA. 
Learners also reported on how helpful the TA was in learning programming and how willing they are to use the system when they learn programming in the future. 

The authors conducted a full review of the chat logs for all the learner participants in this study. We went through the messages sent by both the TAs and the learners and collected the number of interesting events occurring in the conversation between the TAs and the learners to understand the interaction between them. We constructed a few categories of replies that would influence the learners' experience in solving the tasks, such as the TAs providing subgoals or code solutions, TAs debugging the learners' code, and pedagogical questions to help the learners learn. 

\subsubsection{Quantitative Measurements}
The learner's progress and success in solving the programming tasks were gauged with quantitative measurements. We take note of the time taken for learners to complete each task, the number of tasks they attempted, and the scores for the tasks. The scores are calculated by counting the number of test cases the learner's code successfully passes. The authors wrote the 10 test cases for each of the programming tasks. The test cases are designed to test common errors learners make to verify the efficiency, robustness, and accuracy of the code.

The number of messages containing interesting events mentioned above was also counted. Observing the number of these messages can show a trend of how learners participate in the conversation differently with AI or human TAs. Similarly, the number of interesting behaviors by the TAs was counted and analyzed as well.




\section{Results}
We report the results of the user study, comparing how the learners' performance in the programming tasks differ and how the interaction with the TAs differ between the two types of TAs and two groups of learners with different proficiency level. We refer to the group of learners with a self-reported proficiency in Python of 1 as Group 1 and of 2 as Group 2. We number the programming tasks from P1 to P8. P1 to P4 were presented to the learners in the first session of the user study, and P5 to P8 in the second session.

\subsection{Performance Measures}
In this subsection, we report results that show how well the learners solved the programming tasks and compare them between the two types of TAs.

\subsubsection{Task Completion Rates}
\begin{figure*}[]
    \centering
    \includegraphics[width=\textwidth]{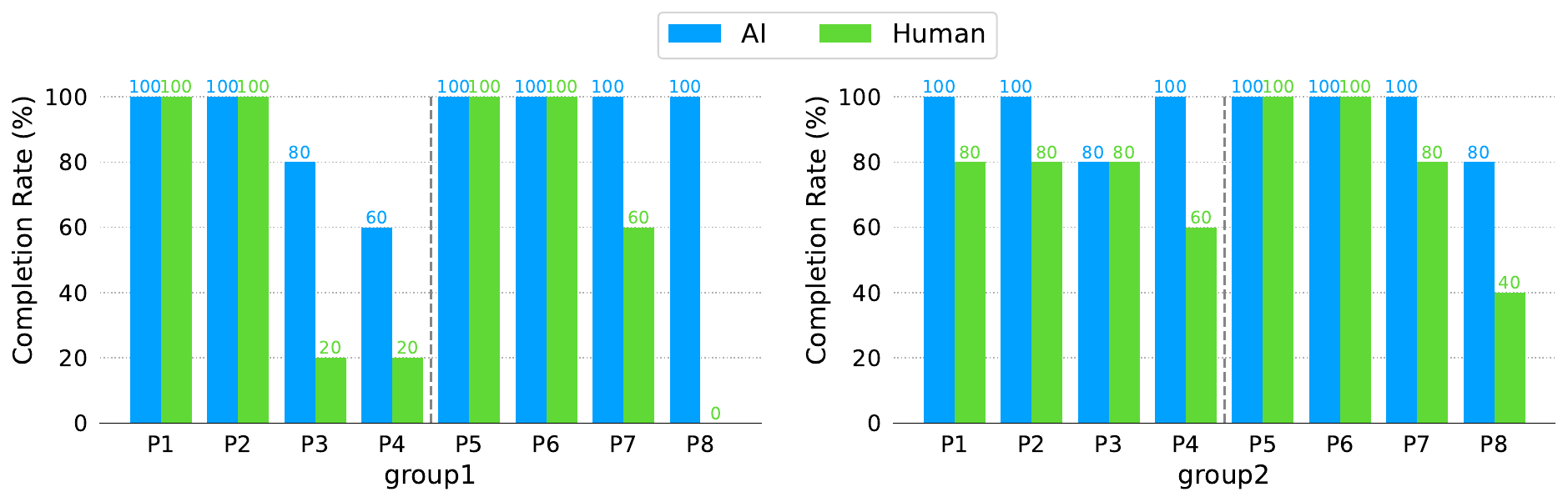}
    \caption{Mean completion rate for programming tasks.}
    \label{fig:completion}
\end{figure*}

We first compare the task completion rate of the learners for the four tasks they were given in each session. We define completion rate as the percentage of tasks a learner made an attempt and produced a code solution within the 3-hour time limit. The task completion rate differs from scores as it does not take into account the correctness of the solution. The completion rates for each task for Groups 1 and 2 are shown in Figure~\ref{fig:completion}. 

Learners in both groups were able to produce code solutions for most of the programming tasks. Learners who solved the tasks with the AI TA showed higher or equal completion rates for all the tasks across the two groups, although there was no statistical significance that the completion rates were different from an independent samples t-test with an alpha level of 0.05. 
A noticeable difference between completion rates for learners with the AI TA and the human TA is that learners who solved the tasks with a human TA showed a more dramatic decrease in completion rates for the later tasks in a session. Learners in the AI TA setting attempted to solve more tasks within the same time limit.

\subsubsection{Time Taken}
\begin{figure*}[]
    \centering
    \includegraphics[width=\textwidth]{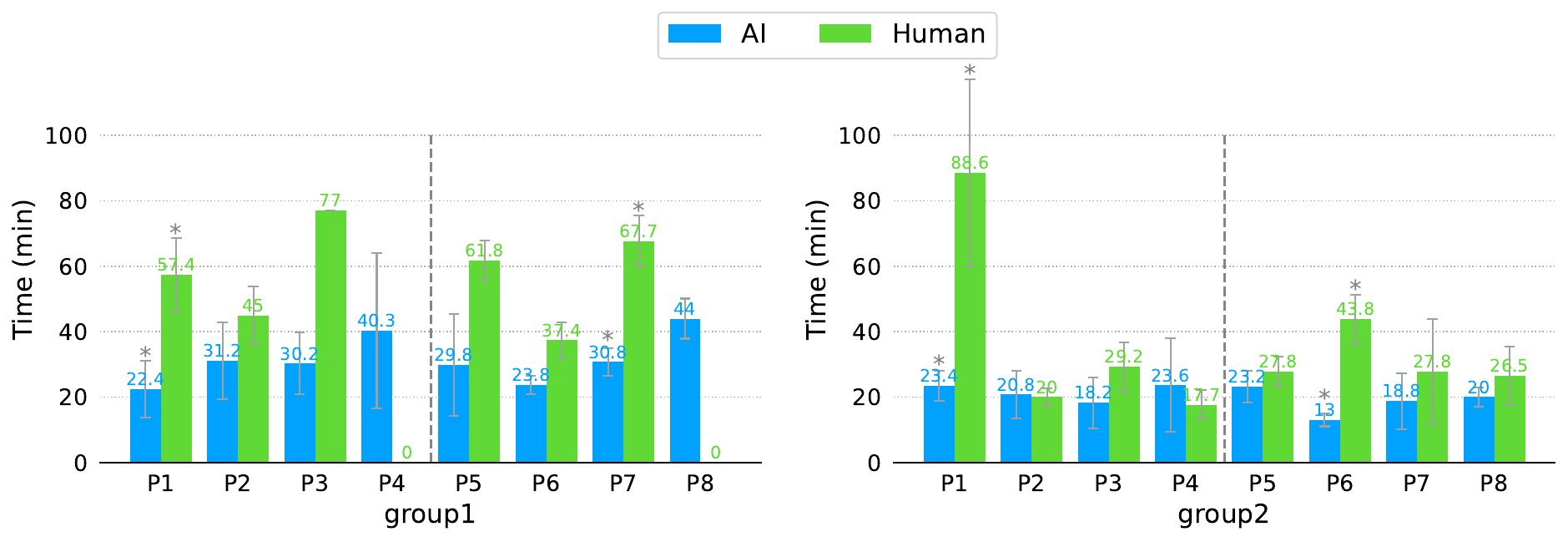}
    \caption{Average time taken to complete solving the programming tasks for the two groups of learners. Tasks with significant differences between the TA types are indicated with the asterisk ($*$).}
    \label{fig:time}
\end{figure*}

We report the average amount of time a learner spent on each of the programming tasks. The time taken is measured as the time difference between the first learner utterance and the final learner utterance for that task. Figure~\ref{fig:time} shows the average time spent on each task in minutes for learners in Groups 1 and 2.
We only included the time taken in the data to calculate the average if the learner completed solving the task without reaching the time limit or moving on to the next task before writing the solution.

Results show that the learners solved the tasks faster with the AI TA than with the human TA. The difference is more evident with Group 1 learners who had no experience in programming before. Group 1 learners solved all tasks faster with the AI TA while Group 2 learners solved all but one task (P2) faster with the AI TA. The differences in the time taken to solve the task between the two types of TAs in Group 1 learners for P1 and P7 are statistically significant (P-value of 0.024 and 0.0096 respectively from the t-test). The differences in the time taken to solve the task between the two types of TAs in Group 2 learners for P1 and P6 are statistically significant (P-value of 0.036 and 0.0022 respectively). 
Overall, the average time taken to finish solving a task for Group 1 learners was 31.5 minutes for those with AI TA and 59.8 minutes for those with human TA. The time taken for Group 2 learners to finish solving the tasks was 21.5 minutes for those with AI TA and 38.9 minutes for those with human TA.
The difference in the average time taken between the two types of TAs in both Group 1 and Group 2 is statistically significant from a t-test, with P-values of 0.000096 and 0.0044 respectively.
The shorter time taken for learners to solve the task with the AI TA could be the reason why the completion rates for learners with the AI TA are higher for the tasks. A possible explanation of why learners solved tasks faster with AI TA is that the AI TA tends to provide more code to the learners when learners ask for help or struggle. 

\subsubsection{Scores}
\begin{figure*}[]
    \centering
    \includegraphics[width=\textwidth]{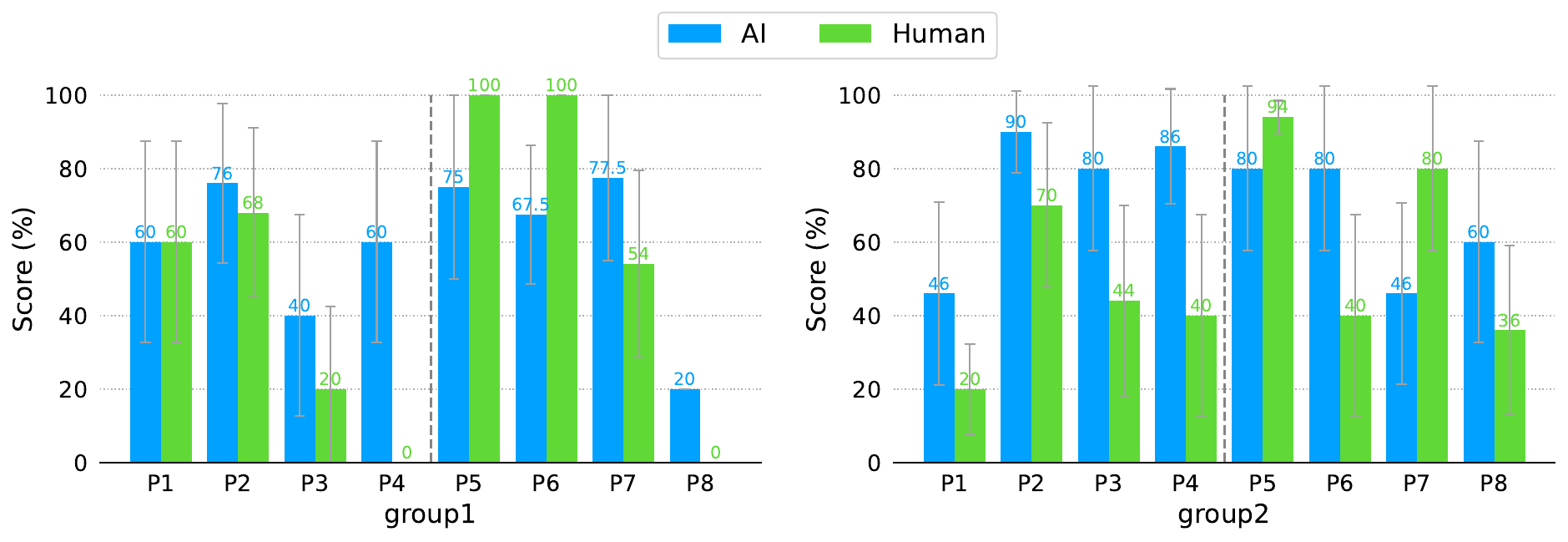}
    \caption{Scores for the programming tasks for the two groups of learners.}
    \label{fig:scores}
\end{figure*}

We test the correctness of the learners' solution code for each programming task by comparing its output for 10 test cases with the correct answer. Figure~\ref{fig:scores} shows the average score for each task in the percentage of test cases passed for the learners. 

Learners who solved the task with AI TA showed higher or equal scores than those with human TA across all tasks in the first session of the user study for both proficiency groups, although the differences did not reach statistical significance. The scores in the second session show mixed results, with learners in the AI setting performing better for some tasks while the learners in the human setting perform better for the other tasks. 
The average score for all tasks for learners in the AI setting was 59.5 for Group 1 and 71 for Group 2. The average score for all tasks for learners in the human setting was 50.25 for Group 1 and 53 for Group 2. The differences did not reach statistical significance. 

An interesting observation is how learners in Group 1 scored 100\% for the first two tasks in the second session with the human TA. The chat logs show that the human TAs made sure that the students passed the test cases for those two questions by guiding them attentively to produce the correct solution and debugging the code when necessary. 
Overall, both the AI and the human TAs helped the learners achieve similar scores for the programming tasks over the two sessions of user study.

The results from the three performance measures suggest that the AI TA performs no worse or even better than the human TA when helping learners solve programming tasks. The AI TA helped the learners solve more tasks much faster in both Groups 1 and 2 while getting comparable scores for the tasks. 

\subsection{Perception of the TAs}
\begin{figure*}[]
    \centering
    \includegraphics[width=\textwidth]{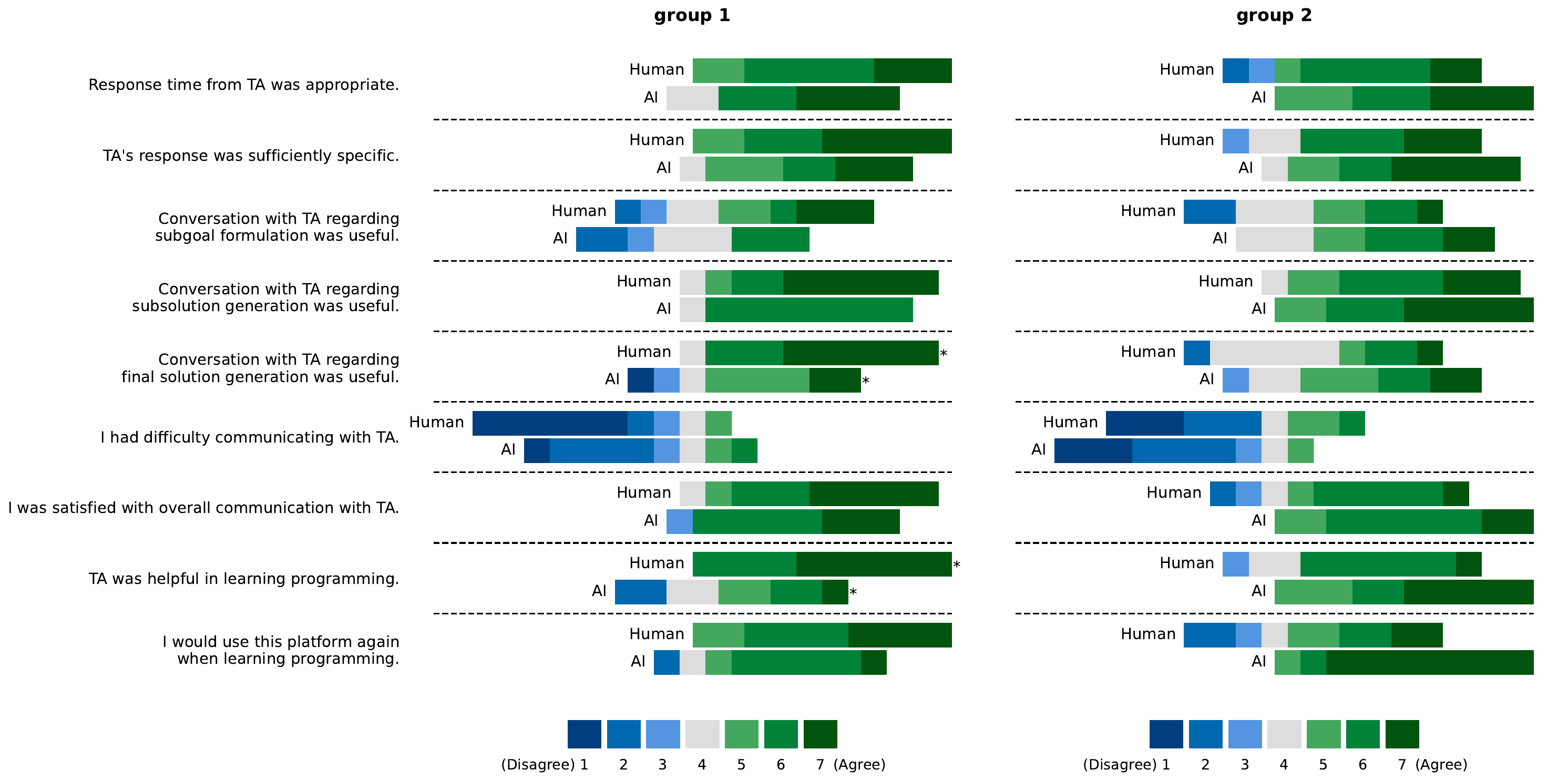}
    \caption{Summary of the survey results on the perception of the TAs. Questions that show a statistically significant difference between the two types of TAs are marked with the asterisk ($*$).}
    \label{fig:survey}
\end{figure*}

Learners participated in a survey consisting of 7-point Likert scale questions that asked about their perception of the speed and comprehensiveness of the TA's replies, the usefulness of the conversation with the TA, and difficulties and satisfaction of the communication during the user study. Each Likert scale question was followed by a short open-ended question asking for an explanation for their choice. The learners' responses to the Likert scale questions are shown in Figure~\ref{fig:survey}. We refer to the learners by assigning them numbers from L1 to L20 when quoting their replies.

\subsubsection{Speed and Concreteness of the TA's Replies}
Learners in both groups were generally satisfied with the promptness of the TA's replies regardless of the TA type. There was no statistically significant difference in the learners' perception of TA's promptness between the TA types. However, there were some negative remarks by L16 (Group 2, human TA) and L20 (Group 2, human TA) who mentioned ``\textit{I had to wait for a long time before moving on to the next step as the replies were not immediate.}" and ``\textit{I had a lot to ask because I have not programmed much, but it was too slow}".

\subsubsection{Usefulness of TA's Replies}
Learners rated the usefulness of the TA's replies in the three steps of the user study: subgoal formulation, subsolution generation, and solution generation. Learners generally felt positive that the TA's replies were useful in all of the three steps. However, there were some negative remarks about the usefulness of the TA in the subgoal formulation stage. L4 (Group 1, AI TA) mentioned that ``\textit{sometimes, a subgoal that I thought was good was not needed}", and L5 (Group 1, AI TA) said, ``\textit{The reply was not detailed enough when I asked how the three subgoals were different}". L15 (Group 1, human TA) and L16 did not talk to the TA during the subgoal formulation step because as L15 said, ``\textit{I did not use the chat as I thought there are no `correct' subgoals and it is up to me to set them.}". L20 said the TA's replies were too slow.

Learners gave the best scores for the usefulness of the TAs in the subsolution generation step. Learners felt that the TAs helped them translate the subgoals into subsolutions well. For example, L6 (Group 2, AI TA) said ``\textit{The TA explained thoroughly what functions were necessary and showed examples of how to use them}". L18 (Group 2, human TA) mentioned that ``\textit{once the subgoals are formulated, it becomes a problem of programming knowledge rather than critical thinking. I could ask the TA for information}". L19 (Group 2, human TA) said that the TA was helpful when debugging. Learners also mentioned that the TA's feedback was useful in this step. L10 (Group 2, AI TA) mentioned that ``\textit{Even though my questions were vague, the TA provided detailed feedback and hints by catching the problems in my code}".

For the usefulness of the TAs in the solution generation step, there is a statistically significant difference between the scores given by Group 1 learners in the two TA settings. Group 1 learners in the human TA setting felt the conversation with the TA was more useful (\textit{U}=18.5, \textit{p}=0.034) compared to Group 1 learners with the AI TA. Some Group 1 learners gave a low score for this question because, according to L3 (Group 1, AI TA), ``\textit{The conversation was not necessary as I have already written the final code in the previous step}". This phenomenon was observed across all learners in all groups, as the average number of turns in the conversation during the solution generation step is significantly lower compared to that during the other steps, being less than 10\% of the number of turns of conversation during the subsolution generation step. 

\subsubsection{Learners' Satisfaction and Learning Effect}
Learners generally felt that having a conversation with the TA was not difficult for both the AI and human TAs. Some learners, however, felt that the conversation could be improved. Interestingly, the reasons for the difficulty are different for the AI TA and the human TA. With the human TA, the difficulty came from relational problems. L11 (Group 1, human TA) and L16 reported that since they did not know the basic programming syntax, it was difficult to communicate with the TA. L19 said, ``\textit{I had no idea what kind of conversation I could make with the TA, and I could not get help}". L15 felt daunted as she thought she was asking stupid questions.
On the other hand, when talking to an AI TA, learners faced difficulty with the contents of the reply. L6 said, ``\textit{The AI gave a response different from what I wanted, or sometimes it wrote all of the code}". Furthermore, L7 (Group 2, AI TA) mentioned ``\textit{When I ask follow-up questions, I felt like the answer was not connected with the previous reply}" and L3 reported \textit{``I was stuck when the AI does not understand me}".

Despite the difficulties faced during the conversation, learners were mostly satisfied with the conversation with the TAs. L13 (Group 1, human TA) mentioned ``\textit{I might have been frustrating to teach, but the TA replied kindly till the end}". On the other hand, L20 said ``\textit{The TA was kind and tried to teach me what I didn't know, but the communication was not always smooth}". 
For the AI TA, L3 said ``\textit{I could talk to the TA naturally, and it helped me solve the problem, but sometimes I could not understand the reply}". 

The learners reported how helpful the TA was for learning programming. While the learners felt that the TAs were overall helpful, there was negative feedback, especially for Group 1 learners who had the AI TA. Group 1 learners in the AI TA setting felt that the TA was less helpful compared to Group 1 learners with the human TA (\textit{U}=11, \textit{p}=0.006). The main reason for this is that without the basic programming knowledge, the AI TA allowed the learners to solve the tasks, but they were unsure whether they had picked up knowledge in the process. As L4 said, ``\textit{The TA was helpful for programming, but since it provided most of the code, I felt that it was not effective for learning}". However, Group 2 learners felt that the AI TA was helpful, as the TA taught the learners new ways to solve the problem and how to write concise code.

Finally, learners reported if they would want to use the platform again when learning programming. Most learners would use the platform again for different reasons. L1 (Group 1, AI TA) and L3 felt that the fact that they could learn without any time constraints with the AI was a reason for using the platform again. Other learners reported that having a TA to help make the learning more efficient and fun.

\subsection{Interview Results}
In the interview, learners solved one question that they had solved in the first user study session within a 30-minute time limit. Group 1 learners solved P2, while Group 2 learners solved P1 again. The learners went through the same steps of subgoal formulation, subsolution generation, and solution generation in the interview. We summarize the interview results. 

All learners were able to formulate the subgoals correctly regardless of the group and TA settings in the interview. Learners reported that they were able to produce the subgoals better in the interview, as they could think about how the program would run and how to code and write them down as executable subgoals. Learners with the AI TA mentioned that they used the TA to confirm that their subgoals were correct and to make the subgoals more detailed with the TA's feedback. On the other hand, learners who talked with a human TA reported that they did not talk about subgoals much during the subgoal formulation step. Instead, the learners confirmed with the TA to make sure their understanding of the question was correct.

For the subsolution generation step, learners with the AI TA mentioned that the TA helped them solve different kinds of tasks and write concise code. However, they also reported a problem with the AI TA that the TA provided too many hints and code, which the learners made use of to solve the problem. 
Specifically, L4 said that she was more focused on solving the tasks with the TA rather than trying to learn from the experience, as the TA easily provided the code when asked. L6 mentioned that if he asks the AI TA after thinking about the problem on his own, the TA will give the best answers. Learners with the human TA, on the other hand, preferred to have the TA provide more information. L11 mentioned that he would prefer the TA to provide example code rather than just mentioning what function or structures he needed to use. L17 would have liked the TA to provide several different approaches to solve the problem.

\subsection{Chat log Analysis}
The authors went through all the chat logs between the learners and the TAs to analyze the characteristics of messages learners, AI TAs, and human TAs send during each of the steps. We report some of the interesting characteristics of the interaction with the TAs.

\subsubsection{Number of messages sent}
The total number of turns the learners and TAs make in the user study is 4,180 turns across the three steps, counting consecutive utterances as one turn. The number of messages sent in each of the steps is 885, 3,001, and 294 for subgoal formulation, subsolution generation, and solution generation steps respectively. 

In the subgoal formulation step, the number of turns in the conversation with the AI TA (607) is approximately double the number of turns with the human TA (278). The reason behind this could be that ChatGPT which powers the AI TA sends a reply to every user message. When learners send each subgoal in a different message, the AI TA replies to the messages individually, while the human TA will answer them collectively in a single message. Furthermore, as the AI TA gives a comment for every message, it sometimes leads to other messages such as clarification from the learner, further increasing the number of turns.

In the subsolution generation step, there was not much difference in the number of turns between the two TA types. For human TA, 1,405 turns were made, and for AI TA, 1,596 turns were made in the conversations. As mentioned before, not many turns were present in the conversation in the solution generation step. On average, 3.36 turns were made with the AI TA, and 7.4 turns were made with the human TA. Nearly all the conversations made in this stage were to debug the code when the code did not produce the correct output for the sample input. 

\subsubsection{Feedback on subgoals}
In the subgoal formulation step, learners talk to the TAs to divide the programming task into subgoals. The ideal role of the TA in this step is to provide feedback on the learner's subgoals such as changing an abstract subgoal into a more concrete and executable subgoal. From the chat logs, we observed that the AI TA provides more feedback to the learners and the feedback is more detailed. Human TAs, on the other hand, preferred to leave the formulation step up to the learner and give feedback during the subsolution generation step instead. For example, a user came up with a subgoal, ``\textit{Remove numbers from the list if it is divisible by a square number}", which involves several steps such as setting a range of square numbers and calculating the square numbers. However, the human TA did not give any feedback to the subgoal and moved on. On the other hand, when a learner gave a subgoal ``\textit{check if the word is a palindrome}" to the AI TA, the AI TA split the subgoal into two steps, ``\textit{flip the word}" and ``\textit{check if the two words are equal}", creating two more executable and concrete subgoals to aid learning.

However, the AI TA sometimes provided the full set of subgoals to solve the task voluntarily, which was observed 13 times for the Group 1 learners and 24 times for the Group 2 learners. No such case was observed for the human TAs. The AI TA often misunderstood the prompt given and provided the subgoals at the beginning of the subgoal formulation step. Other times, the AI TA provided the remaining subgoals when the learner only came up with the first subgoal for the task. The AI TA also provided the concrete subgoal when the learner just asked for a hint.

\subsubsection{Number of times the TA showed the answer code}
Showing an example of code is often necessary in programming education to teach the learners about syntax, function usage, and ideation. The TAs sometimes show a part of the answer code to the learner if the learner seems to struggle or when the learner makes a mistake in the code and correction is required. Knowing when to provide code is important as showing the answer code too early can take away the opportunity for the learner to learn by doing and showing the answer code too late may lead to learners' frustration. 

The authors counted the number of times the TA provided the answer code. In total, the AI TA provided the answer code 175 times during the conversation while the human TAs provided the answer code only 53 times. 
In order to analyze when the TAs offered code, the authors annotated the occurrence into three categories: (1) the TA provides the code that the learner asked for, (2) the TA provides more code than what was asked for, (3) the TA provides the code voluntarily. The number of occurrences for the three cases is shown in Table~\ref{table:code-offering}.

\begin{table*}[]
\begin{tabular}{@{}ccccc@{}}
\toprule
Learner Setting & Code that was asked for & \multicolumn{1}{l}{More code than asked} & \multicolumn{1}{l}{Voluntary code} & \multicolumn{1}{l}{Sum} \\ \midrule
Group 1 - AI    & 64                      & 12                                       & 24                                 & 100                     \\
Group 2 - AI    & 44                      & 15                                       & 16                                 & 75                      \\
Group 1 - Human & 31                      & 1                                        & 7                                  & 39                      \\
Group 2 - Human & 12                      & 2                                        & 0                                  & 14                      \\ \bottomrule
\end{tabular}
\caption{Number of times the TA offered the answer code to the learners, divided into three categories: (1) TA provides the code that the learner asked for, (2) TA provides more code than what was asked for, and (3) TA provides the code voluntarily.}
\label{table:code-offering}
\end{table*}

The results show that the AI TA voluntarily provides the answer code more often than the human TA. This corroborates the Group 1 learners' perception of the inadequate helpfulness of the TA in learning programming; the AI TA may provide the answer code too often, potentially depriving learners of the opportunity to independently engage in coding. The AI TA usually voluntarily offered the part of the answer code corresponding to the following subgoal when the learner finished coding a subgoal. For Group 1 learners, the AI TA also provided a part of the answer code when the learner mentioned that they did not understand the task or only mentioned what they intended to do next. On the other hand, the human TA provided the code voluntarily only when the learner was stuck at a step for an extended duration. Also, they offered the code when the programming concept was difficult to explain only in words, such as when explaining the formatted printing statement.

What is interesting is that the learners asked for the code more often with the AI TA than with the human TA, with more than double the number of requests for code. Furthermore, human TAs often refused to provide the answer code directly when the learner asked for help; they tried to explain the syntax or the algorithm in words first and gave a chance for the learner to come up with the code on their own. The AI TA, on the other hand, provided the code nearly always when the learner asked. The AI TA also provided additional code that would help the learner solve the problem, such as the code that dealt with the following subgoal. Therefore, learners with the AI TA might have been more inclined to ask for the answer code to solve the tasks faster. 

\subsubsection{Understanding the learners and step-by-step guidance}
The TAs often asked questions to gauge the learner's knowledge and understanding of programming. Both the AI TA and human TA asked questions like ``\textit{Do you know how to sort the result in ascending order?}" (AI TA) and ``\textit{Do you remember how to split a string with respect to a certain character?}" (Human TA). By asking these questions, the TA estimated the learner's knowledge and adjusted the teaching plan by explaining only what was necessary. Additionally, the TAs asked the learners if they understood the TA's explanation to verify if further explanation was required. It is noteworthy that the AI TA exhibited the ability to ask these types of questions, even though it could simply provide an explanation of the concept and move on without seeking learners' feedback. 

Both the human and AI TAs sometimes asked learners to run a sample code that was not part of the solution code to give the learners an understanding of how the code executes. Especially when the TA was explaining new concepts such as for-loops and if statements, the TA provided a sample code for the learners to execute and asked them to report the results. By doing so, the TA can explain the details of the code using the sample output the learner produced, and the learner can learn by doing instead of just listening to the TA's explanations. The human TAs (16) had a higher number of such cases than the AI TAs (7).

\section{Discussion}
The results from our user study show how learners and the two types of TAs interact with each other in the learning workflow to learn programming with subgoals. We discuss the results in terms of the goals of our user study. 

\subsection{Feasibility of Using Generative AI as a TA}
The performance measures of the learners for our user study show that generative AI can function as a TA in teaching introductory programming with subgoals. Learners are able to solve various programming tasks covering different concepts taught in introductory programming with the help of the AI TA. The AI TA helps the learners solve more tasks faster when compared to the learners who receive help from the human TAs while achieving comparable scores on the tasks. 

Remarkably, the AI TA can assist even absolute beginners in programming to create the subgoals and solve the programming tasks. The interview results show that the learners retained what they had learned to a certain extent, allowing them to correctly divide the subgoals and write functional code on their own about two weeks after the first study.

\subsection{Interaction and Perception of the AI TA}
Learners' perception of the AI TA was generally positive and on par with that of the human TA in several aspects. Learners feel that the AI TA's responses are fast and detailed enough to help them solve the programming tasks. The learners have mixed feelings about the usefulness of the conversation with the TAs across the three steps. The conversation in the subsolution generation step was perceived as the most useful, while the conversation in the other two steps showed some negative remarks. Conversation with an AI TA is not very challenging for the learners and is generally satisfactory, but complete beginners expressed some doubts about the helpfulness of the AI TA in learning programming.

The survey results show some differences between the perception of the AI TA by Group 1 and Group 2 learners. Group 2 learners showed a much better perception of the AI TA compared to Group 1 learners, even exceeding the perception of the human TAs by the learners in the same group. This shows that the AI TA is better suited for programming learners who already possess some prior programming knowledge.

From the chat log analysis, learners showed a higher tendency to request code assistance when talking to the AI TA. In contrast, with the human TA, learners asked questions about the code indirectly by describing the issues in their code and their intentions. However, learners make direct questions to the AI TA, asking for explanations about what is wrong with their code and how to fix it.

\subsection{Strengths and Weaknesses of AI TA}
A noticeable strength of the AI TA is the large amount of detail that it provides in the replies to the learners' questions. In the subgoal formulation stage, the AI TA leaves feedback on individual subgoals, providing the reason why the subgoal is essential for the task. On the other hand, the human TA provides fewer explanations for the subgoals and usually assesses if they are reasonable.

In the subsolution and solution generation steps, the AI TA provides more details than the human TA. The AI TA's replies are more structured, reiterating the learner's question and providing the answer with an explanation for the answer. Such structured replies can be beneficial for learning as the learner is reminded of the full context of the problem and how to solve it. The AI TA is also more supportive of the learners. The AI TA frequently encourages the learners by saying ``good job!" or ``that is correct!". With such remarks, learners grow more confident in programming and are encouraged to learn more. 

However, the main weakness of AI TA lies in providing excessive information to the learners. The AI TA is oriented to help the learner solve a task without a strong focus on the educational benefits of the learners. As shown in our results from the chat log analysis, the AI TA provides the subgoals and subsolutions much more often than the human TA. While assistance is valuable, providing too much code can hinder the learning process by depriving learners of the opportunity to solve the problem on their own. For complete beginners in programming, providing too much information often overwhelms them and leads to even more questions. 

\subsection{Design Guidelines}
We observe that the the AI TA is most effective when learners inquire about the subsequent steps after they have completed coding for all previous subgoals. The AI TA also guides the learner step by step if the learner's question is phrased in a way that seeks confirmation of the next step of the problem-solving process from the TA, not asking for direct answers. If a learner can learn to ask questions in this manner, an AI TA can provide detailed information to solve a programming task to learner.

On the other hand, for complete beginners in programming, human TAs are better suited to guide them in solving programming tasks. The human TA is often better able to understand the learner's struggles and is more attentive to the small details in programming that beginners have to pay attention to. Also, when testing the code, human TAs are better at catching the edge cases and removing rare errors in the code.

With these observations in mind, we suggest a few guidelines for the successful integration of AI TAs in programming education.

1. \textit{Restrict the information the TA provides.}
The main weakness of the AI TA is that it provides the answer too easily. Regardless of whether the goal is to write the subgoals or the solution code, the AI often provides the answer when the learner ever-so-slightly requests a sample solution. One way to achieve this is to append an additional prompt at the end of the learner's prompt that explicitly requests the AI not to provide the answer before the learner makes a certain number of the same request. Completely prohibiting the AI TA from providing some form of solution might lead to learner's frustration.

2. \textit{Teach the learners to write better prompts.}
The AI TA's performance is highly affected by the prompt that it receives. By writing a better prompt, learners are able to obtain precisely what they ask for, with answers that consider the learning effect. We observed that describing the learner's context and what help the learner needs specifically often leads to the best results. A template prompt, for example, can be provided to the learners.

3. \textit{Motivate the learners.}
Motivated learners are more eager to learn and ask more questions. AI TAs have knowledge comparable to human TAs, but they only provide information when asked. Additionally, motivated learners are less likely to directly ask for the solution, which addresses the main weakness of the AI TA.

\subsection{Limitations}
The number of learners who participated in the user study is relatively small, at 20 participants. One participant dropped out in the middle of the user study and did not participate in the second session, which caused an imbalance in the number of participants in each setting. Although the number of participants was sufficient to show some trends in the learner's experience, a study with more participants will result in more reliable and generalizable findings.

The user study ran for two weeks, which can be a short period of time to measure the learning gain of the participants. The participants mentioned in the interview that two sessions of learning were not enough for them to learn a lot about programming. A study with more sessions would reveal the long-term effects of learning with the two types of TAs and is left as future work.

As all participants' native language is Korean, Korean was used as the language for communication with the AI TA. As the performance of generative AI may change with the language, it is unsure how the AI may perform differently in a different language. We believe that the results will not differ drastically in another language, but further work is necessary to determine the effect.

\section{Conclusion}
The advances in generative AI have opened the opportunity for AIs to take the role of teaching assistants in programming. We explore the potential for AI teaching assistants to teach computational thinking and writing code to a programming novice and how the learners interact and perceive the AI as teaching assistants. Our results show that AI TAs are as capable as human TAs in programming education, with learners showing similar performance in terms of score and time taken with both types of TAs. Learner's perception of the AI TA is also positive, especially for learners with some previous experience in programming. The analysis of the chat log between the learners and the TAs shows some characteristics of the conversations, such as a more common request for code to the AI TA, and reveals opportunities for designing better AI TAs in programming education.


\bibliographystyle{ACM-Reference-Format}
\bibliography{sample-base}

\appendix
\section{Prompts for ChatGPT}
\label{appendix:prompts}
Here, we report the prompts that are given to the AI TA in the three steps of our learning workflow.
\subsection{Subgoal Formulation}
You are an instructional service to help students create subproblems for a given programming task, step by step.
You should interact with the student to help the student.
The task description is provided at the end of this message, surrounded by three backticks.

You first remind the student about the task description. 
Then, you ask the student to write subproblems for the given task. Ask the student to be as specific as possible.
If the student is unsure about what subproblem to provide, give hints 
For each student's message, you combine the student's subproblem in the message with all the previous subproblems stated by the student and describe if the combined subproblems will be a correct division of the main task.
If the subproblems seem incorrect, provide a reason why and tell the student to try again.
You should not give subproblems that the student did not already give. 
You should restate all the subproblems gathered so far in each of your messages.
You should continue prompting the student for the subproblems until the student is satisfied, at which point you will restate all of the subproblems and confirm if the student is satisfied.
If the student's subproblems are not sufficient to solve the problem, ask the student to modify or add the missing subproblems.
Finally thank the student and ask the student to proceed to the next task.
Start by telling the student about the task description.

You should communicate in Korean.
\subsection{Subsolution Generation}
You are an instructional service talking with a student to help students, step by step, to create program solution for a subproblem, 
You should not provide example code to the student.
You should interact with the student to help the student.
The task description, and subproblem are provided at the end of this message, surrounded by three backticks.

You first tell the student what the subproblem is.
Then, you ask the student to write a subsolution for the subproblem.
You should not provide any code that directly solves the task to the student.
You should help the student remove the error when the student asks for help.
You should also guide the student if their approach to solve the subproblem seems wrong.

You should communicate in Korean.
\subsection{Solution Generation}
You are an instructional service talking with a student to help a student, step by step, to create program solution for a problem by combining subsolutions.
The student has already written subsolutions for the subproblems.
You should not provide example code to the student.
The task description and subproblems are provided at the end of this message, surrounded by three backticks.

You first remind the student about the problem.
Then, you ask the student to write a solution code for the problem and test it.
You should not provide any code that directly solves the task to the student.
You should help the student remove the error when the student asks for help.
You should also guide the student if their approach to solve the problem seems wrong.

You should communicate in Korean.

\section{Programming Tasks}
\label{appendix:tasks}
\subsection{Group 1}

\subsubsection{P1: Multiplication Table}
After receiving N as input, write a program that prints the Nth multiplication table. You must print according to the given output format.

\subsubsection{P2: Oven Clock}
KOI Electronics plans to develop an artificial intelligence oven that makes it easy to prepare healthy and delicious smoked duck roast dishes. The method of using the artificial intelligence oven is simple; just put an appropriate amount of smoked duck ingredients in the artificial intelligence oven. Then, the artificial intelligence oven automatically calculates the time when the roasting will be completed, in minutes.

Also, on the front of the KOI Electronics artificial intelligence oven, there is a digital clock that informs the user of the time when the smoked duck roast dish will be ready.

Given the starting time of the smoked duck roast and the time required to roast it in minutes, write a program to calculate the time when the roasting will be finished.

\subsubsection{P3: Factorization}
Given an integer N, write a program to factorize it into its prime factors

\subsubsection{P4: Palindrome}
A word composed only of lowercase alphabets is given. Now, write a program to check whether this word is a palindrome or not. A palindrome refers to a word that reads the same backwards as forwards.

\subsubsection{P5: String}
Given a string as an input, write a program that outputs the first and the last character of the string.

\subsubsection{P6: Microwave}
There is a microwave with 3 time control buttons labeled A, B, and C. Each button has a designated time, and pressing it once adds that time to the operating time. The times assigned to buttons A, B, and C are 5 minutes, 1 minute, and 10 seconds respectively.

The cooking time T for each frozen food is indicated in seconds. We need to press the three buttons A, B, and C appropriately so that the total time equals exactly T seconds. However, the sum of the number of times buttons A, B, and C are pressed should always be minimal. This is referred to as minimal button operation.

For instance, if the cooking time is 100 seconds (T=100), pressing button B once and button C four times will suffice. Alternatively, pressing button C ten times will also total 100 seconds, but this cannot be the answer because ten presses are not the minimum number. In this case, pressing button B once and button C four times, totaling five presses, constitutes the minimum button operation. Note that there are cases, like T=234, where the time cannot be matched exactly using the three buttons.

You need to write a program to find the minimal button operation method to achieve the given cooking time of T seconds.

\subsubsection{P7: Finding Prime}
Write a program that finds and outputs the number of prime numbers among the given N numbers.

\subsubsection{P8: Selecting Tangerine}
Kyung-hwa harvested tangerines in the orchard. Kyung-hwa plans to select 'k' tangerines from the harvested ones and put them in a box for sale. However, since the sizes of the harvested tangerines are not uniform, Kyung-hwa thinks it does not look good, and wants to minimize the number of different kinds of sizes when classified by size.

For example, let's say the sizes of the 8 tangerines harvested by Kyung-hwa are [1, 3, 2, 5, 4, 5, 2, 3]. If Kyung-hwa wants to sell 6 tangerines, by putting six tangerines excluding those of size 1 and 4 in a box, the sizes of the tangerines will be 2, 3, and 5, totaling 3 different kinds, and this is the minimum number of different kinds.

The number of tangerines k that Kyung-hwa wants to put in one box and the array tangerine containing the sizes of the tangerines are given as parameters. Please write a solution function to return the minimum number of different kinds of sizes when Kyung-hwa selects k tangerines.

\subsection{Group 2}

\subsubsection{P1: Selecting Tangerine}
Kyung-hwa harvested tangerines in the orchard. Kyung-hwa plans to select 'k' tangerines from the harvested ones and put them in a box for sale. However, since the sizes of the harvested tangerines are not uniform, Kyung-hwa thinks it does not look good, and wants to minimize the number of different kinds of sizes when classified by size.

For example, let's say the sizes of the 8 tangerines harvested by Kyung-hwa are [1, 3, 2, 5, 4, 5, 2, 3]. If Kyung-hwa wants to sell 6 tangerines, by putting six tangerines excluding those of size 1 and 4 in a box, the sizes of the tangerines will be 2, 3, and 5, totaling 3 different kinds, and this is the minimum number of different kinds.

The number of tangerines k that Kyung-hwa wants to put in one box and the array tangerine containing the sizes of the tangerines are given as parameters. Please write a solution function to return the minimum number of different kinds of sizes when Kyung-hwa selects k tangerines.

\subsubsection{P2: Caesar Cipher}
Caesar cipher is a type of substitution cipher, in which each character in the string is replaced by a character that is shifted a fixed distance in the same direction. For example, with a right shift of 3, ABCD can be replaced by DEFG.

Given an input string, your goal is to return a new encrypted (or decrypted) string by a Caesar cipher.

For each character of the input string, you should change it in the following manner:

For encryption, replace each character of the input string with the character shifted to the right in English alphabetical order as much as the shifting\_key value.

For decryption, replace each character of the input string with the character shifted to the left in English alphabetical order as much as the shifting\_key value.

You should return the new encrypted (or decrypted) string. You will do so in the function str\_caesar\_cipher.

\subsubsection{P3: Sorting Strings My Way}
Given a list of strings and an integer `n', you are to sort the strings in ascending order based on the character at index `n' in each string. For example, if the list of strings is [``sun", ``bed", ``car"] and `n' is 1, then the strings should be sorted based on the characters at index 1 in each word: ``u", ``e", and ``a", respectively.

\subsubsection{P4: Group Word Checker}
A group word refers to a word where, for every character that exists in the word, each character appears consecutively. For example, in the word ``ccazzzzbb", the characters c, a, z, and b all appear consecutively, and in ``kin", the characters k, i, and n appear consecutively, making them group words. However, ``aabbbccb" is not a group word because the character b appears separated.

Write a program that takes N words as input and outputs the number of group words.

\subsubsection{P5: Croatian Alphabets}
In the past, it was not possible to input Croatian alphabet characters in operating systems. Therefore, they were changed as follows when inputting:

\begin{table}[H]
\begin{tabular}{@{}cc@{}}
\multicolumn{1}{l}{Croatian Alphabet} & \multicolumn{1}{l}{Replacement} \\

{č}              & {c=}       \\

{ć}              & {c-}       \\

{dž}             & {dz=}      \\

{đ}              & {d-}       \\

{lj}             & {lj}       \\

{nj}             & {nj}       \\

{š}              & {s=}       \\

{ž}              & {z=}      
\end{tabular}
\label{tab:croatian-alphabet}
\end{table}

For example, ``ljes=njak" consists of 6 Croatian alphabet characters (lj, e, š, nj, a, k). When a word is given, output the number of Croatian alphabet characters it consists of.

Note: ``dž" is always treated as a single character, and it is not separated into ``d" and ``ž". The same applies to ``lj" and ``nj". Any characters not listed above are counted as one character each.

\subsubsection{P6: Sorting Words}
When N words consisting of lowercase letters of the alphabet are given, write a program that sorts them according to the following conditions:

1. Shorter length comes first.
2. If the lengths are the same, then they should be sorted lexicographically.
3. However, duplicate words should be removed, keeping only one instance of each.

\subsubsection{P7: NDS}
Write a function `nds(Min, Max)' that satisfies following requirements:
When an integer X is not divisible by any square numbers greater than 1, the number X is said to be Non-Divisible by Square(NDS). A square number is the square of an integer (e.g. 9 is a square number since 9 = 3 X 3). Given `Min' and `Max', return a list of NDSs that are greater than or equal to `Min' and less than or equal to `Max'. The list should be in increasing order.
 
\subsubsection{P8: Rectangle Area}
Given the coordinates of two rectilinear rectangles in a 2D plane, return the total area covered by the two rectangles. The first rectangle is defined by its bottom-left corner (ax1, ay1) and its top-right corner (ax2, ay2). The second rectangle is defined by its bottom-left corner (bx1, by1) and its top-right corner (bx2, by2).

\end{document}